\documentclass[a4paper,fleqn,usenatbib]{mnras}

\usepackage[T1]{fontenc}
\usepackage{ae,aecompl}
\usepackage{epsfig}
\usepackage{extarrows}
\usepackage{graphicx}	% Including figure files
\usepackage{amsmath}	% Advanced maths commands
\usepackage{amssymb}	% Extra maths symbols

\usepackage{colortbl}
\definecolor{mygray}{gray}{.9}
\definecolor{mygray1}{gray}{.8}
\definecolor{mygray2}{gray}{.7}

\usepackage[noend]{algpseudocode}
\usepackage{algorithmicx,algorithm}

\usepackage{palatino}
\usepackage{tikz}
\usetikzlibrary{shapes.geometric, arrows}
\usepackage{color}

\usepackage{ulem}

\title[ ]{Pulsars Detection by Machine Learning with Very Few Features}

\author[H.T. Lin et al.]{
Haitao Lin$^{1}$,
Xiangru Li$^{2}$\thanks{Correspondence author's E-mail: xiangru.li@gmain.com (X.Li)},
Ziying Luo$^{1}$
\\
$^{1}$School of Mathematical Sciences, South China Normal University,  Guangzhou 510631, China\\
$^{2}$School of Computer Sciences, South China Normal University,  Guangzhou 510631, China\\
}

\date{Accepted 2020 January 20. Received 2020 January 19; in original form 2019 May 16}

\pubyear{2020}

\begin{document}
\label{firstpage}
\pagerange{\pageref{firstpage}--\pageref{lastpage}}
\maketitle

\begin{abstract}

It is an active topic  to investigate the schemes based on machine learning (ML) methods for detecting pulsars as the data volume growing exponentially in modern surveys.
To improve the detection performance, input features into an ML model should be investigated specifically.
In the existing pulsar detection researches based on ML methods, there are mainly two kinds of feature designs: the empirical features and statistical features. Due to the combinational effects from multiple features, however, there exist some redundancies and even irrelevant components in the available features, which can reduce the accuracy of a pulsar detection model. Therefore, it is essential to select a subset of relevant features from a set of available candidate features and known as {\itshape feature selection.} In this work, two feature selection algorithms ----\textit{Grid Search} (GS) and \textit{Recursive Feature Elimination} (RFE)---- are proposed to improve the detection performance by removing the redundant and irrelevant features. The algorithms were evaluated on the Southern High Time Resolution University survey (HTRU-S) with five pulsar detection models.
The experimental results verify the effectiveness and efficiency of our proposed feature selection algorithms. By the GS, a model with only two features reach a recall rate as high as 99\% and a false positive rate (FPR) as low as 0.65\%; By the RFE, another model with only three features achieves a recall rate 99\% and an FPR of 0.16\% in pulsar candidates classification. Furthermore, this work investigated the number of features required as well as the misclassified pulsars by our models.

\end{abstract}


%%%%%%%%%%%%%%%%%%%%%%%%%%%%%%%%%%%%%%%%%%%%%%%%%%%
%
\section{Introduction}
Since the discovery of the first pulsar in 1967 by Jocelyn Bell Burnell and Antony Hewish \citep{Hewish 1968}, many researchers and scholars in astronomy and astrophysics devote themselves to pulsar discovery, detection and application such as interstellar navigation \citep{Coll 2009}, indirect detection of gravitational wave \citep{Taylor 1994} and potential discovery of dark matter \citep{ Baghram 2011}. As of September 2019, the Australia Telescope National Facility (ATNF)  \citep{manchester2005parkes} has identified 2796 pulsars. However, it is estimated that there are 140,000 pulsars undiscovered in our galaxy \citep{zhang 2016}. Therefore, various surveys have been carried out and expected to discover other pulsars.
The new and improved radio telescopes such as the Square Kilometre Array \citep[SKA; ][]{Smits:2009} and
Five-hundred-meter Aperture Spherical radio Telescope \citep[FAST; ][]{Smits:2009b,NanRD:2006,NanRD:2011,NanRD:2016}, will bring about rapidly growing data volumes and pulsar candidate numbers. In these candidates, only a small number of them are real pulsars, while {the overwhelming majority of them are from various forms of noises or radio frequency interference (RFI). To find the real pulsars buried in the complex noises, one important step is to pick out worthwhile candidates. These selected candidates are of higher probabilities to be signals from real pulsars, and deserve more time and resources for some follow-up observations and further identifications. This
procedure is known as \textit{candidate selection}.

With the arrival of the big data era of pulsars,
candidate-selection methods  have been extended to machine learning (ML).
 For instance, \citet{Eatough et al. 2010} used artificial neural networks \citep[ANN;][]{Bishop 2006, Hastie 2001} in Parkes Multibeam Pulsar Survey  \citep[PMPS;][]{manchester2001parkes}. {\citet{Morello 2014} developed an ANN approach, Straightforward Pulsar Identification using Neural Networks (SPINN),  for pulsar candidate classification in the HTRU survey.  \citet{lyon 2016} proposed a learning classifier called  Gaussian Hellinger Very Fast Decision Tree \citep[GHVFDT;][]{Lyon 2014} to deal with the real-time streams of pulsar candidates.
And \citet{Zhu 2014} introduced a scheme called Pulsar Image-based Classification System (PICS) by combining algorithms such as ANN, Support Vector Machine \citep[SVM;][]{Cortes and Vapnik 1995} and convolutional neural network \citep[CNN;][]{LeCun:1998, Bengio:2009}. More studies based on the machine learning methods for pulsar candidate classifications can be found in \citet{Bates:2012}, \citet{Bethapudi 2018}, \citet{Tan 2017} and \citet{Guo 2019}.

For an ML model, the design of its inputs is essential. These inputs are referred to the features of pulsar candidates. There are many approaches to extract and construct pulsar features from surveys. Generally, there are two kinds of features of pulsar candidates. One is called empirical features such as the dispersion measure (DM), the period and the signal to noise ratio (SNR). The other one is called statistical features such as the mean or the standard deviation of the folded profile. However, not all features are necessary for an ML model as some of them may be redundant or ineffective. This is one typical challenge in choosing proper inputs to an ML-based pulsar detection model.

This work investigated the feature selection problem for classifying the pulsar candidates using supervised machine learning methods \citep{Mitchell 1997}. The aim is to improve the pulsar candidate classification performance by removing the redundant features. This process of reducing features is called \textbf{feature selection} or \textbf{variable selection}. To be specific, feature selection is to choose a subset of features as the input of a  ML model.

One of our motivations is that the researches on  feature selection methods are inadequate in pulsar candidate classification literatures. As we know, feature selection plays an indispensable role in a model and the performance of feature selection should be evaluated when new features were proposed before modeling. The importance of features were usually measured by mutual information \citep{Shannon:1948}, Pearson product-moment correlation coefficient or Relief-based algorithms \citep{Kononenk 1994}. However, these feature evaluations weren't used for automatically selecting features. Besides, all these available feature evaluations are filter-based approaches, which completely on the relation between the features and its labels, and assume that these features are independent of each other.
In applications, however, it is inevitable that there exist some dependencies and redundancies in our obtained features. These redundancies and correlations in the input features can be harmful to the performance of a model \citep{Yu 2004}.
Thus, some new algorithms of feature selection should be designed to solve these problem.

Another motivation of feature selection comes from its potential benefits: making data visualized and understandable, reducing the measurement and storage requirements, reducing training and prediction time, defying the curse of dimensionality to improve prediction performance \citep{Guyon 2003}.
In existing literatures of ML-based pulsar detection scheme, the number of the features for models is more than six, and some even up to twenty-two, which means one should collect enough information and extract a lot of features before training a model. If we are able to  predict a candidate correctly only by a small number of features, it makes candidate selection less complex, since we just need to focus on these very few but relevant features, instead of all the features.

In our work, two wrapped methods \citep{Kohavi 1997} of feature selection were implemented to minimize the number of inputs. It is a catch-all group of techniques with performing feature selection as part of the model construction process, involving a kind of supervised ML called {\itshape{ensemble learning} \citep{Opitz:1999}.
To evaluate the performance of our work, these ensemble learning models were trained and tested,
with the number of features varying from one to four. In Section 2, several ML concepts are introduced, including the data set,  sampling types,  ensemble models and the performance measures. In Section 3, 18 candidate features used for this study are introduced. In Section 4, two feature selection algorithms, \textit{Grid Search} (GS) and \textit{Recursive Feature Elimination} (RFE), are proposed, where univariate-feature selection, double-features selection and multiple-feature selection are implemented and analyzed. Further discussions on the number of features of a model, the feature candidate pool and the misclassified pulsars are presented in Section 5. Conclusions of this work are made in Section 6. %In addition, a suspected pulsar was discovered in our experiment.

%%-----------------------------------------------------------------------------------------------------
\section{Machine Learning Concepts}\label{sec:MLConcepts}

This work investigated the pulsar candidate classification problem based on some supervised machine learning methods \citep{Mitchell 1997}.
A supervised machine learning method associates a set of labelled data, denoted as $D=(X,Y)$, where $X$ represents the feature(s) set of the data and $Y$ the label set.
The idea of a supervised ML model is to learn a function $f$ mapping $X$ to $Y$.
A perfect model for data $D$ will make $f(X)=Y$. In real applications, however, it is hard to obtain a perfect mapping. Thus, we try to find an approximation, $\hat{f}$, for the function $f$ by minimizing a loss function which measures the distance between the real labels $Y$ and the prediction $\hat{f}(X)$. This will result in a model that predicts most labels correctly, which in our case pulsars and non-pulsars.

In order to utilize an ML scheme, the following fundamental considerations should be made: a) Sampling methods; b) Selection of ML models; c) Candidate features and features selection; d) Performance measures of models. We will discuss a), b), d) in Sections \ref{sec:MLConcepts:DataPreparation}, \ref{sec:MLConcepts:EnsembelLearning} and \ref{sec:MLConcepts:PerformanceMeasure} respectively. The procedure c) will be discussed in Sections \ref{sec:Features} and \ref{sec:FeatureSelection}, which is the main work discussed in this paper.

\subsection{{Data preparation}}\label{sec:MLConcepts:DataPreparation}
\subsubsection{{Candidate generation}}

Pulsar candidates from the Southern High Time Resolution university survey \citep[HTRU-S;][]{Keith 2010} were investigated in our work. The HTRU-S is an ongoing all-sky
survey for radio pulsars at Parkes radio telescope in Australia and processed by GPU-based PEASOUP pipeline \citep{Keith 2010}.
The dataset is public\footnote{https://drive.google.com/drive/folders/0B3-6QHgPuBInTkNzLWREX3ZzUTg} \citep{Morello 2014}, consists of 91192 XML files, each of which describes the information of a pulsar candidate with a label.

To generate these pulsar candidates, a number of procedures were involved in dealing with the original signals from the receiver of a radio telescope.
Generally, RFI excision is done in the first step by clipping the strong interference \citep{Keith 2010}.
Next, dedispersion is conducted to eliminate the influence of the dispersion by trying various DM values. Dispersion is a phenomenon that signal components with lower frequencies arrive later than those with higher frequencies, due to free electrons between the pulsar and Earth, and DM value is the amount of dispersive delay  \citep{Lorimer 2009}.
Then, fast Fourier transform (FFT) is adopted to identify the possible period of signals \citep{Lorimer 2012}.
Also, initial acceleration searches \citep{Johnston 1991} are done on the Fourier transformed data, which looks for signals that have variation in period over time.
After a periodic signal is found in a dedispersed time-series, the raw data is folded to the DM and period of the signal. The folded candidate data cube is then optimized by gird searching through the Period-DM space and shifting the data cube in time and frequency to find the values that give the highest profile SNR. The acceleration curve in the diagnostic plot is also generated similarly by shifting the data cube by
considering the effect of acceleration which varies the period over time.
Finally,
by summing the transformed data cube over both the time and the frequency, the folded profile of a candidate is obtained.

% the sub-bands plot and the sub-integrations plot are obtained by summing the data cube over the time and over the frequency, respectively. And the folded profile of the candidate by summing all the frequency channels and time inter

This process outputs pulsar candidates with both a number of physical and statistical values and six diagnostic plots. Fig. \ref{pulsarfig} shows some information on a pulsar candidate. On the left, the plots from top to bottom  are : the sub-bands plot (marked with \textbf{A} in  Fig. \ref{pulsarfig}) which records the pulse in different bands of observed frequencies, the sub-integrations  plot (\textbf{B}) which records the pulse in the time domain, and the folded profile (\textbf{C}) which is the folded signal from the pulsar that was collected over the entire observation. On the right, the plots from top to bottom  are : the grid searching plot (\textbf{D}), the DM searching curve (\textbf{E}), and the accelerating searching curve (\textbf{F}). Besides, several physical values such as the Bary period and the best DM are summarized.

Ideally, a promising pulsar candidate has the following characteristics in its diagnostic plots (Fig. \ref{pulsarfig}):
the signal is sufficiently strong in the folded profile (\textbf{C} in Fig. \ref{pulsarfig}); a persistent pulse signal is observed in both time and frequency domain (\textbf{B} and \textbf{A} in Fig. \ref{pulsarfig}, correspondingly);  it peaks at a non-zero DM (\textbf{E} in Fig. \ref{pulsarfig}); there is a strong peak in the grid search plot of period vs. DM (\textbf{D} in Fig. \ref{pulsarfig}).
However, diagnostic plots may not always be obvious and there are exceptions to
almost every rule. Sometimes, real pulsars can have characteristics that we usually associate
with RFI, and vice versa. In this case, follow-up observations will be needed.

\begin{figure}
\centering
\includegraphics[width=8cm]{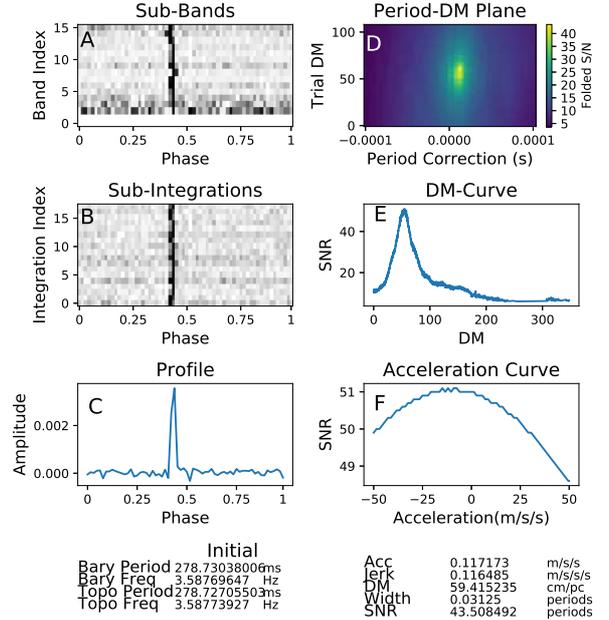}
\caption{Candidate information of XML file pulsar\_0033.phcx on HTRU-S, showing six diagnostic plots and summary statistics. More details are in Section 2.1.1. }\label{pulsarfig}
\end{figure}

%-----------------------------------------------------------------------------------------------------
\subsubsection{{Dealing with Class Unbalance}}\label{sec:MLConcepts:DealingUnbalance}

%-----------------------------------------------------------------------------------------------------
The HTRU-S consists of 1196 pulsars
as well as 89996 non-pulsar candidates,
which is highly unbalanced. Candidates labelled as pulsars are much fewer than  those labelled as non-pulsars   with a ratio of 1:75. The model trained in an unbalanced dataset tends to judge an unknown candidate as a non-pulsar instead of a pulsar "for safe". According to the experience in \citet{Lyon 2013}, recall rates consistently dropped as the class imbalance increased.
To alleviate this bias, we used three sampling techniques to balance the training data, i.e., random under-sampling, random over-sampling and Synthetic Minority Over-sampling Technique \citep[{SMOTE};][]{Chawla 2002}.
These three sampling techniques were explained as follows.
Suppose that $T$ is a two-class training data set with $P$ its minority class set and $N$ its the majority class set, i.e., $T = P \cup N$.
 The under-sampling tries to balance $T$ by randomly removing samples from its majority class set $N$, while the random over-sampling alleviates the class imbalance by randomly choosing samples with replacement from its minority class set $P$.
The process of SMOTE involves k-Nearest Neighbors \citep[KNN;][]{Altman 1992}.
To create a synthetic sample of minority class, a random sample $p$ was taken from the minority class set $P$ firstly, and then $k$ nearest neighbors of $p$ can be found in $P$, denoted as $\{p_{i}\}_{i=1}^{k}$. Next, $p'$ in $\{p_{i}\}_{i=1}^{k} $ is randomly chosen, and finally we got a synthetic sample by selecting a random point on the line segment which joins $p$ and $p'$.
It can be expressed by $\lambda p + (1-\lambda) p'$, where $\lambda$ is a random number between 0 and 1. This process is repeated until $T$ is balanced.
SMOTE has been used in pulsar searching and pulsar classification \citep{Devine 2016, Bethapudi 2018}.
%Our implementation currently uses five nearest neighbors.

The sampling process of our experiments designed as follows.
Firstly, the data was randomly separated into two subsets, a training set and a testing set with 50\% of the data each with 598 pulsars and 44998 non-pulsars in each subset.
Then, three kinds of samplings techniques: under-sampling, over-sampling and SMOTE, were implemented to the training data to get three balanced training data sets, while the test set was kept unchanged in every trial (Table \ref{sample}).
We will compare the performance of the models with different sampling types in Section \ref{sec:FeatureSelection}.

\begin{table}
        \begin{center}
        \caption {The description of the dataset obtained via different sampling methods. 50\% of all the candidates are used as training data which is denoted as {"NonSampling"}, while others are in a test set. The balanced training dataset generated by under-sampling, over-sampling and SMOTE are denoted as "UnderSample", "OverSample" and "SMOTE", correspondingly.}\label{sample}
        \begin{tabular}{lllllll}
        \hline \noalign{\smallskip}
         Data &Sampling & Pulsar & Non-pulsar & Total & Ratio   \\
         \noalign{\smallskip}
        \hline\noalign{\smallskip}
         Train&NonSampling  &598&{44998} & {45596} &1:75\\
        &SMOTE  & {44998}& {44998} & {89996} &1:1\\
        &UnderSample & 598 & 598& 1196 &1:1\\
        &OverSample  & {44998}& {44998} & {89996} &1:1\\
        \noalign{\smallskip}
         Test&  &598&{44998} & {45596} &1:75\\
          \noalign{\smallskip}
        Total&  &1196&{89996} & {91192} &1:75\\
         \noalign{\smallskip}
        \hline
        \end{tabular}
        \end{center}
\end{table}

\subsection{Ensemble Learning Algorithms}\label{sec:MLConcepts:EnsembelLearning}
Five models are used in our work (Table \ref{ourmodel}). All of them are relevant to a large category of ML called \textbf{ensemble learning} \citep{Opitz:1999}.
Ensemble learning is a machine learning paradigm where multiple learners are trained to solve the same problem. An ensemble algorithm contains a number of learners which are usually called base learners (or weak learners). {Base learners are usually \textbf{Decision Trees} (Section 2.2.1) . Ensemble methods try to construct a set of base learners and {aggregate their classification results}, which usually perform better than a single base learner \citep{zhou:2016}.

The reason why we use the ensemble learning methods instead of other ML models is that it can be less sensitive to the over-fitting problem than some other learning classifiers (e.g., ANN) and sometimes it is even able to reduce the generalization error after the training error has already reached zero.
{Usually, the generalization ability of an individual learner is much weaker than a model of ensemble learning. Three reasons can be found by viewing the nature of ML as searching a hypothesis space for the most accurate hypothesis \citep{Dietterich 1997}.
Firstly,  the training data might not provide sufficient information for choosing a single best learner.
Secondly, the search processes of the learning algorithms might be imperfect. For example, even if there exists a unique best hypothesis, it might be difficult to achieve this optimal one due to our limitation on computational resources and computational time. Fortunately, the ensemble methods can compensate for such imperfect search processes. Thirdly, the hypothesis space being searched might not contain the true target function, while the ensemble method can give some good approximation. For example, it is well-known that the classification boundaries of decision trees are linear segments parallel to coordinate axes. If the target classification boundary is a smooth curve, it is difficult to obtain a good result using a single decision tree; however, it is possible to compute a good approximation by combining a set of DTs.

Ensemble approach consists of three representative methods:  Boosting \citep{Schapire:1990}, Bagging \citep{Breiman:1996} and Stacking \citep{Wolpert:1992}.
In this work, we considered a single classifier (DT) and other four ensemble learning methods (Table \ref{ourmodel}).
The various algorithms used in this work
are described in the following subsections.
\begin{table}
        \begin{center}
        \caption {Models in our work. }\label{ourmodel}
        \begin{tabular}{lllllll}
        \hline \noalign{\smallskip}
          Model &Abbr.&Type \\
        \hline\noalign{\smallskip}
          {Classification And Regression Tree} & {CART}      &DT \\
          {Adaptive boosting }                 & {AdaBoost } &Boosting    \\
          {Gradient Boosting Classifier }      & {GBoost}    &Boosting    \\
          {eXtreme Gradient Boosting }         & {XGBoost}   &Boosting    \\
          Random Forest                        & RF          &Bagging   \\

         \hline\noalign{\smallskip}

        \end{tabular}
        \end{center}
\end{table}

\subsubsection{Decision Tree}

\textbf{Decision Tree} \citep[DT;][]{Quinlan:1986} is a popular tool in ML and usually chosen as a base learner in ensemble methods. A DT consists of a
set of tree-structured decisions working
in a divide-and-conquer way. A bifurcation of a tree is called a node, while a leaf node is the end of a branch of a DT. Each non-leaf node, also called a split point, is associated with a classification based on one feature. In detail, {the} data falling into the split point will be split into multiple subsets according to their values on the feature that best split the data. Each leaf node is associated with a label, which will be assigned to the instances falling into this node. In prediction, a series of feature tests are conducted starting from the root node, and the result is obtained when a leaf node is reached. In our experiment, we used the \textbf{ Classification And Regression Tree} \citep[CART;][]{Breiman 1984}.
The CART algorithm generates a DT using a \textbf{Gini coefficient}, which measures the impurity of the data and defined as follows.
$$Gini (D) = 1-\sum_{k=1}^{J} p_k^2, $$
where $D$ represents a set of training instances with $J$ classes, $p_k$ is the proportion of training instances from the $k$-th class.
Each feature $a$ in $D$ is associated with a Gini coefficient which is defined as
$$Gini(D,a) = \sum_{v=1}^{V} \frac{|D^v|}{|D|} Gini(D^v) ,$$
where $\{D^v\}_{v=1}^{V}$ is a partition of $D$ divided by feature $a$, and $Gini(D^v)$ is the Gini coefficient in the subset $D^v$. The feature with the minimum $Gini(D,a)$ will be used in a split point.
\subsubsection{Boosting {Algorithms}}
The concept of boosting was first proposed by \citet{Schapire:1990}. It aims to generate a strong classifier by combining a series of learning weak classifiers by weighting training data.
Three common boosting algorithms are Adaptive boosting \citep[Adaboost;][]{Freund 1995}, Gradient Boosting Classifier \citep[GBoost;][]{M 2016} and eXtreme Gradient Boosting \citep[XGBoost;][]{Chen 2016}.

The \textbf{AdaBoost} is implemented by sequentially learning a series of weak learners and adaptively adjust the weights of the samples based on the performance of the weak learners. A weak base learner is trained by training samples with the same weight. These weights are related to the weak learners' accuracy. Actually, these weak learners are trained sequentially, and the weight of every sample is also correspondingly adjusted after a weak learner being trained. Subsequent weak learners pay more attention to those instances misclassified by previous learners. The outputs of all weak learners are combined into a weighted sum that represents the final output of the boosted learner. Compared with the traditional AdaBoost, the process of \textbf{GBoost} is to construct a series of weak learners  by using a gradient descent method and combining them to be a strong classifier. The \textbf{XGBoost} algorithm is an upgraded version of GBoost. In XGBoost, the loss function is added with some regularization techniques in case of over-fitting. Further details of the various boosting algorithms are available in the references above.

\subsubsection{Random Forests}
 \textbf{Random Forests} \citep[RF;][]{Barandiaran:1998} is an ensemble learning algorithm by constructing a large number of DTs {for classification and regression.
 The main process of RF is as follows. Step 1, randomly sample a data set from our training data set with replacement. This data set {had} the same size as our training data and {was} called a bootstrap set.
Step 2, choose a subset of the available features randomly to fit a DT on this bootstrap set.
 {Then, repeat Step 1-2 to construct a series of different DTs.
 Finally, these DTs {were} combined together so as to form a RF whose output {was} the average results of each DT. RF has been proved to be an effective algorithm to solve over-fitting \citep{Breiman 2001}.

\subsection{Performance Measure} \label{sec:MLConcepts:PerformanceMeasure}

\begin{table}
        \begin{center}
        \caption {{Performance measures and their definition.} } \label{meansure}
        \begin{tabular}{lllc}
        \hline\hline\noalign{\smallskip}
         {Measure Name} &{Definition}\\
                 \noalign{\smallskip}
        \hline\noalign{\smallskip}
        {Accuracy}&     {$\frac{TP+TF}{TP+TF+FP+FN}$}\\
                \noalign{\smallskip}
        {Recall}  &     {$\frac{TP}{TP+FN}$}\\
                \noalign{\smallskip}
        {Precision}&    {$\frac{TP}{TP+FP}$}\\
                \noalign{\smallskip}

        {$F_1$-score}&  {$\frac{2\times Precision \times Recall}{Precision +Recall}$}\\
                \noalign{\smallskip}

         {FPR}& {$\frac{FP}{TN+FP}$}\\
        \noalign{\smallskip}\hline
        \end{tabular}
        \begin{tabular}{lllc}
        {Note:}& \\
        {TP}& {Num of pulsars correctly classified} \\
        {TN}& {Num of non-pulsars correctly classified}\\
        {FP}& {Num of non-pulsars incorrectly classified as pulsars}\\
        {FN}& {Num of pulsars incorrectly classified as non-pulsars}\\
        \end{tabular}
        \end{center}
\end{table}

To measure the performance of a classifier, some metrics were defined ( Table \ref{meansure}). Accuracy is a ratio of correctly predicted candidates to the total candidates.
In many real-world cases, accuracy {is} not the best quality metrics for classification. In fact, our data is highly imbalanced. The noise and RFI actually represent the overwhelming majority of the data points. In this case, pulsars often are} misjudged in the ML models but the accuracy may {be} still very high. Thus accuracy is not helpful to measure the performance of our models.

Instead, we utilized recall, precision, $F_1$ score and false positive rate (FPR), to measure the performance of our models. Recall is the ratio of pulsars correctly predicted to all the pulsar signals. A perfect recall rate means we find all the pulsars in our test set. Precision is the fraction of the pulsars correctly predicted among all instances which are predicted to be pulsars.
 However, as recall rate increasing the precision can be decreasing, which means more non-pulsars are misjudged. To trade off these two measures, \textbf{$F_1$ score} is introduced, which is the harmonic mean of the recall and precision.
 Also, we used FPR to measure the ratio of misjudged non-pulsars to the total number of non-pulsars.

Generally, recall and FPR are two primary metrics and widely used to measure the quality of classifiers of pulsar detection. In pattern recognition tasks with binary classification (pulsars and non-pulsars),  recall measures how many pulsars could be correctly predicted; FPR measures how many
non-pulsars would be predicted incorrectly as pulsars. The former shows the ability and reliability of {detecting} the real pulsars. The closer to 1, the better. The latter measures the misjudgement of non-pulsars as pulsars. The greater, the larger cost of the verification work in the later stage.
Thus, our algorithms of feature selection {were} designed to maximize the recall as the first objective  and to minimize the FPR as the second objective.

\section{Features of Pulsar Candidates}\label{sec:Features}
In this section, we first summarized the features used in machine learning models for pulsar detection , and then described our pulsar candidate features.

\subsection{Overview of Pulsar Features}

When using ML to candidate selection,
the design of features is one of the most important steps before training a classifier.
Features are required to be distinguishable between pulsars and non-pulsars.
The first application of ML in pulsar candidate selection
\citep{Eatough et al. 2010} designed 12 features to construct ANNs.
These 12 features were extracted from a graphical selection software \texttt{JREAPER} \citep{Keith 2009}, including 2 from pulse profile, 3 from DM curve, 3 from acceleration curve, 2 from sub-band plot and 2 from sub-integration plot. However, the model was
only able to detect $\sim$50\% of the pulsars with periods below 10 ms.  To deal with this limitation, \cite{Bates:2012} used another ANN with as many as 22 features, most of which were developed as an advancement of the works by \cite{Keith 2009} and \cite{Eatough et al. 2010}. These features consisted of 4 basic parameters (SNR, DM, period and width of pulse) and 18 others extracted from the folded profile, DM curve and frequency sub-band data. It is shown that their model with these 22 features was capable of detecting pulsars at all pulse periods, but was appreciably less adept at identifying strong candidates with a large pulse duty cycle or with millisecond periods.
After then, \cite{Morello 2014} proposed six empirical features (ID 1-6 in Table \ref{fea1to22}) to develop a system named \textbf{SPINN} which obtains excellent performance.
These features followed three guidelines to maximize the classification performance, i.e., they were required to, firstly, reduce selection effects against faint or more exotic pulsars, secondly, be robustness to noisy data, and thirdly, be relevant and limited in quantity.

Instead of empirical features, \cite{lyon 2016} attempted to train an ML model with 8 statistical features (ID 7-14 in Table \ref{fea1to22}), including the mean, the standard deviation, the kurtosis and the skewness of  both the folded profile and DM curve, respectively. Based on these features, an algorithm called \textbf{Gaussian Hellinger Very Fast Decision Tree} \citep [GHVFDT;][]{Lyon 2014} were built for survey-independent applications and continuous data-stream processing. However, \cite{Tan 2017}  found that it was hard to detect those pulsars with wide integrated pulse profiles using the GHVFDT with these eight features. Thus, additional features were proposed to complement those by \cite{lyon 2016}, which involved computing the correlation coefficient between each sub-band and the folded profile, as well as between each sub-integration and the folded profile.
Recently, image features have been used to construct the deep neural nets as the data processing capability of modern computers develop. \cite{Zhu 2014} had worked on the pulsar classification problem with a system \textbf{Pulsar Image-based Classification System} (PICS), which was a process of image pattern recognition of deep neural nets and whose inputs were
four image features, including the folded profile, time versus phase plot, frequency versus phase plot and DM curve. All of the image features were computed using a software called \texttt{PRESTO}  \citep{ Ransom 2011}. Other features for ML models can be referred to \cite{Mohamed 2018} and \cite{Guo 2019}.

\subsection{Our Candidate Features}

Considering the performance of the existing ML-based works \citep{Bates:2012, Morello 2014, lyon 2016} carried on HTRU-S, we used eighteen candidate features in our work (Table \ref{fea1to22}). Among them, six empirical features (ID 1-6) are from \citet{Morello 2014};  Eight features (ID 7-14)
    were defined by \citet{lyon 2016}; The remaining four features (ID 15-18) were extracted statistically from the acceleration curve. The notations and the definitions of these features are described as follows.

\begin{enumerate}
\item [(1)] {\itshape{Log of the signal-to-noise ratio of the folded profile} ($\log(\texttt{SNR})$)}. SNR is a measure of the purity of the signal. In our work, we took the logarithm form of SNR, $i.e., \log(\frac{1}{\sigma \sqrt{w}}\sum_{p_i\in W}(p_{i}-\bar{b}))$, where $p_i$, $w$ are respectively the amplitude and width of folded profile in pulse region $W$, and $\bar{b}$, $\sigma$ are respectively the mean and the standard deviation in the off-pulse region.

\item  [(2)]{\itshape{ Intrinsic equivalent duty cycle of the profile}} \cite[$D_{eq}$;][] {Morello 2014}.
    This measure is computed  based on the duty cycle of pulse by removing the dispersive smearing time $\Delta \tau$ \cite {Morello 2014}: $D_{eq} = \frac{w-\Delta\tau}{P}$, where $w$, $P$ are  the width and period of a candidate in seconds, respectively.

\item  [(3)]{\itshape{ The logarithm of the radio of barycentric period and dispersion measure }} ($Log(P/DM)$).  This  feature fuses the information of $P$ and $DM$, and it was found helpful to categorize pulsar candidates into RFI, general pulsars and millisecond pulsars.

\item  [(4)]{\itshape{ Validity of optimized dispersion measure}} ($V_{DM}$). $DM$ of a pulsar exhibits a non-zero value while $DM$ of RFI is usually measured to be close to zero. A feature is defined as $\tanh(DM-DM_{min})$ to separate them, where $DM_{min}$ is the minimum of $DM$ of pulsars in HTRU-S.

\item  [(5)]{\itshape{Persistence of signal through the time-domain}} ($\chi_{(SNR)}$) .  A real pulsar is expected to emit pulse signal during the observation time. To measure the signal persistence, $\chi_{(SNR)}$ is defined as the average scores of $\chi_{(s)}$, where $\chi_{(s)}$ represents the "score" of each sub-integration, i.e.,
                                $
                                {\chi_{(s)}=}
                                \begin{cases}
                                1-exp(-\frac{s}{b}),  &s\geq 0,\\
                                \frac{s}{b},  &s<0,
                                \end{cases}$
     where $s$ is the SNR in a sub-integration, and $b$ is a threshold.

\item  [(6)]{\itshape{ {Root-mean-square (RMS)} distance between the folded profile and the sub-integrations}} ($D_{RMS}$). For RFI components, they probably show some phase drift and some unstabilities.
$D_{RMS}$ is defined as $\sqrt{\frac{1}{w n_{sub}}\sum_{i\in W}\sum_{j}(p_i-s_{ij})^2}$ to describe the stability of the signals from the observations, where $p_i$ is the value of the $i$th bin in the folded profile, $s_{ij}$ is the value of $i$th bin and $j$th sub-integration, $n_{sub}$ is the number of sub-integration.

\item  [(7) -(10)] The mean { ($Pf_{.mean}$)}, standard deviation { ($Pf_{.sd}$)}, kurtosis { ($Pf_{.kurto}$)}, skewness { ($Pf_{.skew}$)} of folded profile.
 These features are defined as follows:
\begin{equation}\label{eq_mean}
{Pf_{.mean} = \operatorname{\mu}= \mathrm{E}\left[p \right]}{,}
\end{equation}
\begin{equation}\label{eq_sd}
{{Pf}_{.sd} = \operatorname{\sigma} = \mathrm{E}\left[\left(p-\mu\right)^{2} \right]}{,}
\end{equation}
\begin{equation}\label{eq_kurto}
{Pf_{.kurto}=\mathrm{E}\left[\left(\frac{p-\mu}{\sigma}\right)^{4}\right]=\frac{\mathrm{E}\left[(p-\mu)^{4}\right]}{\left(\mathrm{E}\left[(p-\mu)^{2}\right]\right)^{2}}}{,}
\end{equation}
\begin{equation}\label{eq_skew}
{Pf_{.skew}=\mathrm{E}\left[\left(\frac{p-\mu}{\sigma}\right)^{3}\right]=\frac{\mathrm{E}\left[(p-\mu)^{3}\right]}{\left(\mathrm{E}\left[(p-\mu)^{2}\right]\right)^{3 / 2}}}{,}
\end{equation}
where $p$ is a variable of the folded profile, and $\mathrm{E}\left[. \right]$ denotes the expectation of a variable.

\item [(11) -(14)]  The mean ($DM_{.mean}$), standard deviation { ($DM_{.sd}$)}, kurtosis { ($DM_{.kurto}$)}, skewness { ($DM_{.skew}$)} of the dispersion measure curve. The definitions of these four statistics on dispersion measure are  given by {replacing} $p$ in {Equations} \eqref{eq_mean}- \eqref{eq_skew} with {the} variable \texttt{DM}.

\item[(15) -(18)]  The mean ($Ac_{.mean}$), standard deviation ($Ac_{.sd}$), kurtosis ($Ac_{.kurto}$), skewness ($Ac_{.skew}$) of the acceleration curve.
    These four features are extracted from the acceleration curve, similar to those based on folded profile and \texttt{DM}.
\end{enumerate}

%------------------------------------------------------------------------------------------------------
\begin{table}
\begin{center}
        \caption {Notations and definitions of the eighteen candidate features. }\label{fea1to22}
    \begin{tabular}{llllllllll}
        \hline
        \noalign{\smallskip}
         ID &Feature &Description \\
        \noalign{\smallskip}
        \hline\noalign{\smallskip}
        1&  $log(\texttt{SNR})$& log of signal-to-noise of the pulse profile\\
        \noalign{\smallskip}
        2&$D_{eq}$& Intrinsic equivalent duty cycle of the               \\
                 && pulse profile  \\
        \noalign{\smallskip}
        3&$Log(P/DM)$&      log of the ratio between period and DM        \\
        \noalign{\smallskip}
        4&$V_{DM}$&       Validity of optimized DM                        \\
        \noalign{\smallskip}
        5&$\chi_{(SNR)}$  &   Persistence of the signal in the time domain\\
        6&$D_{RMS}$& RMS between folded profile and                        \\
        &&sub-integration       \\
        \noalign{\smallskip}
        7&  $Pf_{.mean}$ & Mean of the folded profile                  \\
        8&$Pf_{.sd}$   & Standard deviation  of the folded profile                   \\
        9&$Pf_{.kurto}$ & Kurtosis of the folded profile                  \\
        10&$Pf_{.skew}$ & Skewness of the folded profile                   \\
        11&  $DM_{.mean}$ & Mean of the DM curve                 \\
        12&$DM_{.sd}$   & Standard deviation of DM curve                  \\
        13&$DM_{.kurto}$ & Kurtosis of DM curve                  \\
        14&$DM_{.skew}$ & Skewness of  DM curve                  \\
        15&$Ac_{.mean}$ & Mean of the acceleration curve                 \\
        16&$Ac_{.sd}$   & Standard deviation of acceleration curve                  \\
        17&$Ac_{.kurto}$ & Kurtosis of acceleration curve                 \\
        18&$Ac_{.skew}$ & Skewness of acceleration curve                  \\
        \noalign{\smallskip}\hline
    \end{tabular}
\end{center}
\end{table}

\section{Feature selection}\label{sec:FeatureSelection}
We have prepared eighteen features as the candidate features in the previous section and in this section, \textbf{feature selection} will be made.

Then, two main missions of this} work are to determine the appropriate number of features for pulsar candidate classification and select the features.
These problems will be investigated in the following two subsections and each subsection consists of two stages: the algorithm and the result.
\subsection{Univariate-Feature Selection \& Double-Feature Selection }
Firstly, we considered the case of feature selection with only one or two features for the models and evaluate their performances. It is expected to be impossible to detect pulsars {accurately} from massive candidates by only one or two features since the distributions of pulsars and non-pulsars overlap on every single feature.
However, the motivation to do this is to identify one or two of the most decisive features for pulsar candidate classification and study whether some discoveries can be made based on these few features.

\subsubsection{Grid Search (GS) Algorithm}
The idea of the GS algorithm for feature selection arises from grid search.
It can produce a globally optimal result to the problem but with a substantial computational cost. In our work, the GS is proposed as follows: Firstly, we consider each possible subset $S_1$, where $S_1$ is a subset of $S$ and the number of its elements is a preset integer $len$. Next, the recall rate is calculated for a Model $\mathfrak{L}$ using the features in each this kind subset $S_1$ and we filter out most of the subsets by a recall threshold $\tau$.
Since  the objective of GS is to first increase the
recall rate and then decrease the FPR, we set $\tau$ {close to 1} in this step ($\tau$ = 0.95 in our trials). Finally, we calculate the subset $S^*$ as the selected features  with the minimal FPR.  If $len = 1$, the output is the result of  univariate-feature selection. If  $len = 2$, the output is the result of  double-feature selection. The GS algorithm is presented in Algorithm \ref{algorithm1}.

\begin{algorithm}
\caption{Grid Search (GS)} \label{algorithm1}
\hspace*{0.02in} {\bf Input:}\\
\hspace*{0.12in} $D$;  \,\quad \quad \quad \quad\quad\quad \quad  A set of training data\\
\hspace*{0.12in} $S$;   \,\ \quad \quad \quad \quad\quad\quad \quad  A set of feature candidates  \\
\hspace*{0.12in} $\mathfrak{L}$;    \,\ \quad \quad \quad \quad\quad\quad \quad Model \\
\hspace*{0.12in} $Len$;    \,\,\quad \quad \quad\quad\quad \quad  The reset number of selected features\\
\hspace*{0.12in} $\tau$; \ \, \quad \quad \quad \quad\quad\quad \quad  Recall threshold\\
\hspace*{0.02in} {\bf Output:}\\
\hspace*{0.12in} $S^*$.      \quad \quad \quad \quad \quad \quad \quad A set of  selected  features from $S$\\
\begin{algorithmic}[1]
\For {$S_1 \subset S$ and $|S_1|= Len$}
\State $S_{\tau} =  \{S_1 |Recall(S_1;\mathfrak{L}, D)\ge \tau\}$
\EndFor

\If {$S_{\tau} \neq \emptyset$}
\State $S^* = \arg \min_{S_1\in S_{\tau}} \text{FPR}(S_1;\mathfrak{L}, D)$
\Else
\State $S^*= \arg \max_{S_1 \subset S,|S_1|= Len} {Recall}(S_1;\mathfrak{L}, D)$
\EndIf
\State Return $S^*$
\end{algorithmic}
\end{algorithm}

\subsubsection{Results}\label{sec:FeatureSelection:result1}

GS was implemented on our models except for RF.  RF was not used in this case since there were not enough features for it to generate various random DTs. The experimental details are described in Section \ref{sec:FeatureSelection:implemantation}.

The results of selected features and the metrics of model performance were presented in Table \ref{single_double_result}.
For example, in case of using DT as a classifier and being trained with unbalanced data, the selected feature is ID 10 for univariate-feature selection, resulting in {a recall of 80\%}, a precision rate of 94\%, an $F_1$ score of 86\% and an FPR of 0.062\%. In another case, the best features are ID 1 and ID 10 for DT trained with SMOTE data set, resulting in {a recall of 98\%}, a precision rate of 68\%, an $F_1$ score of 80\% and an FPR of 0.062\%.
The selected features depend on the {classifiers used} as well as the sampling types. We made the following observations:

\begin{table*}
    \caption{ The results of univariate-feature selection (on the left of each table cell) \& double-feature selection (on the right of each table cell).
    Based on the GS, the performance metrics of our models were calculated with four types of sampling, including ordinary training set (noted as "NonSampling"), SMOTE training set ("SMOTE"), under-sampling training set ("UnderSample") and over-sampling training set ("OverSample"). Selected features depend on the used classifiers as well as the sampling types. More analyses is presented in Section \ref{sec:FeatureSelection:result1}.%It shows that feature of ID 10 is the most frequently chosen in both univariate-feature and double-feature selection. Besides, most models with single/double features achieved either low recall or high FPR. Among them, AdaBoost was best whose selected features are ID 1 and ID 10 and achieved a recall rate of 0.99 and an FPR of 0.651\%.
    } \label{single_double_result}
\begin{tabular}{lllllllllllllllll}
\hline
Method&Sampling&feature ID&Recall {(\%)}&Precision {(\%)}&$F_1$ Score  {(\%)}&FPR (\%)\\
\hline
DT        &NonSampling&10/3,10&{80/91}&{94/93}&{86/92}&{0.062/0.093}\\
          \rowcolor{mygray}
          &SMOTE&{5}/1,10&{{98}/98}&{{32}/68}&{{47}/80}&{2.718/0.062}\\
          \rowcolor{mygray1}
          &UnderSample&{2}/1,9&{{93}/96}&{{31}/46}&{{45}/{70}}&{2.796/1.040}\\
          \rowcolor{mygray2}
          &OverSample&{2}/2,10&{{92}/95}&{{41}/80}&{{56}/87}&{2.784/0.307}\\
AdaBoost   &NonSampling&10/5,10&{78/90           }&{97/93}&{86/91}&{0.003/0.096}\\
          \rowcolor{mygray}
          &SMOTE&{10}/1,10&{ {93} /99}&{{34}/67}&{{49}/80}&{2.433/0.651}\\
          \rowcolor{mygray1}
          &UnderSample&{2}/1,10&{{93}/99 }&{{31}/58}&{{47}/73}&{2.738/0.962}\\
          \rowcolor{mygray2}
          &OverSample&{2}/2,10&{93/97       }&{34/64}&{50/77}&{2.405/0.733}\\

XGBoost&NonSampling    &10/2,10&{78/91}&{94/93}&{86/92}&{0.064/0.056}\\
          \rowcolor{mygray}
          &SMOTE&{10}/1,10&{{93}/98}&{{34}/68}&{{49}/80}&{2.433/0.622}\\
          \rowcolor{mygray1}
          &UnderSample&5/1,10&{98/99}&{30/59}&{46/74}&{3.007/0.925}\\
          \rowcolor{mygray2}
          &OverSample&{10}/6,10&{{91}/95}&{{45}/82}&{{60}/88}&{1.469/0.278}\\
GBoost    &NonSampling&10/2,10&{79/91}&{13/93}&{22/92}&{6.958/0.096}\\
          \rowcolor{mygray}
          &SMOTE&5/1,10&{98/98}&{32/69}&{48/81}&{2.762/0.593}\\
          \rowcolor{mygray1}
          &UnderSample&{5}/1,10&{{95}/99}&{{27}/58}&{{42}/73}&{3.429/0.964}\\
          \rowcolor{mygray2}
          &OverSample&{10}/2,10&{{90}/94}&{{25}/89}&{{48}/91}&{3.762/0.156}\\

\hline
\end{tabular}
\end{table*}

\begin{figure}
  \centering
  \includegraphics[width=10cm]{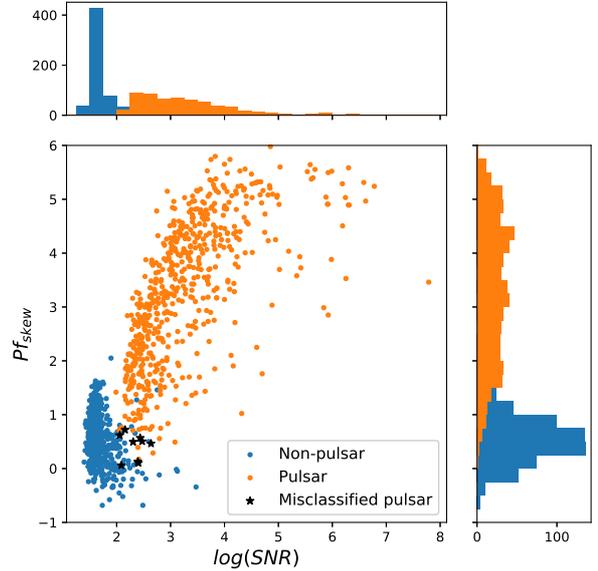}\\
  \caption{ The distribution of the under-sampling dataset in the  $log(\texttt{SNR})$  (ID 1) vs. $Pf_{skew}$ (ID 10) space. Most pulsars have positive $Pf_{skew}$ and higher SNR and it is shown that most of the non-pulsars can be easily separated from pulsars. The misclassified pulsars were collected from boosting models based on the double features, and most of these misclassified ones lie near the boundary between pulsars and non-pulsars.} \label{1to10}
\end{figure}

\begin{itemize}
\item \textit{{The influence of class imbalance.} }Models trained using the unbalanced set are better on precision but worse on recall in the cases using single and double variables, while models trained on the balanced set are better on recall rate and lower on precision rate and consequently a large FPR. These results coincide with the conclusion in \cite{Lyon 2013}.
\item \textit{{Performance of the models.}} Most models based on single/double features achieved either low recall or high FPR. Models with double features are better than those with a single one. Often, an additional feature provides more useful information for the classifier to distinguish the pulsars from non-pulsars, and thus the classifier achieves a better performance.
\item {\textit{Selected features.}}  {Feature with} ID 10 ($Pf._{skew}$) is the most frequently selected in both the univariate-feature and the double-feature selection. Besides, by double-feature selection, the model with only two features can achieve presentable performance.
    It is shown that AdaBoost with features of ID 1 and ID 10, can achieve {99\%} on recall and 0.651\% on FPR. To illustrate the separability of these two selected features, the distribution of some candidates on the SNR (ID 1) vs. $Pf_{skew}$ (ID 10) space was given in Fig. \ref{1to10}. it is shown that most of the non-pulsars can be easily separated from pulsars based on these two features.
     In fact, the GS algorithm is a wrapped method which tries every possible combination of feature candidates, can obtain a globally optimal solution to  double-feature selection. However, the importance of each feature in a model can not be measured using this method.

\item {\textit{DT vs. ensemble algorithm.}} The performance of the ensemble algorithms is as good (bad) as a single DT algorithm after univariate-feature or double-feature selection.
    The reason is that the very few features limit the ability of ensemble learning models.

\item {\textit{Misclassified pulsars.} } The recall is as high as {99\%} for most of the models after double-feature selection, which implies that only a few pulsars were misclassified. The pulsars misclassified by at least two of the boosting models were plotted in Figure \ref{1to10}, which shows that they are mostly near the decision boundary between pulsars and non-pulsars.
\end{itemize}

An obvious fact is that the performance of the models based on double-feature selection were better when compared with those based on univariate-feature selection. Thus, to improve the performance, a reasonable intuition is to explore the cases with more selected features.

\subsection{Multiple-Feature Selection}
Although the GS algorithm also can select multiple features under the necessity of our problem, its computational burden are heavy in this kind of cases.
In this subsection we hence designed another feature selection algorithm (see Algorithm \ref{algorithm2}) called \textbf{Recursive Feature Elimination} (RFE), which eliminates the worst feature recursively based on the measure recall.
The GS and RFE will be compared with each other on computational complexity \citep{Papadimitriou 2003} in Section \ref{sec:ComputationComlexity}.

\subsubsection{Recursive Feature Elimination Algorithm}

The idea of RFE is to eliminate the worst feature in a recursive way. The ``worst" refers to the feature with the lowest contribution to a model in the available features.
The objective function of RFE is to maximize the recall rate.
 Here, $S$ is the set of the feature candidates and $\mathfrak{L}$ a model. A feature $s^*$ with the least contributions to the model $\mathfrak{L}$ will be eliminated from $S$. The ready-to-eliminate feature $s^*$  means that the model based on the
feature set $S\backslash \{s^*\}$ results in the maximum recall rate: $s^* = \arg \max_{s\in S} Recall(S\backslash \{s\})$. Denote $S^* = S\backslash \{s^*\}$. Then, repeat the process of elimination until the number of the elements in the $S^*$ reaches an} expected number (three or four in this work).
 In essence, RFE is a greedy algorithm to maximize the recall rate when the features were eliminated one by one.
\begin{algorithm}
\caption{Recursive Feature Elimination} \label{algorithm2}
\hspace*{0.02in} {\bf Input:}\\
\hspace*{0.12in} $D$;\quad  \quad \quad \quad \quad \quad \quad {\footnotesize A set of training data}\\
\hspace*{0.12in} $S$;  \quad  \quad \quad \quad \quad \quad \quad {\footnotesize A set of feature candidates}  \\
\hspace*{0.12in} $\mathfrak{L}$; \quad  \quad \quad \quad \quad \quad \quad {\footnotesize Model}\\
\hspace*{0.12in} $Len$;    \quad \quad \quad\quad\quad \quad  {\footnotesize The expected number of selected features}\\
\hspace*{0.02in} {\bf Output:} \\
\hspace*{0.12in} $S^*$.       \ \  \quad \quad \quad \quad \quad \quad {\footnotesize The set of selected features from $S$}\\
\begin{algorithmic}[1]
\State $S^* = S $   \  \quad \quad \quad \quad
\While { $len(S^*)$>$Len$ }
\For {$s \in S$ } \quad \quad \quad \quad \quad \quad \quad \quad
\State $R(s) = Recall(S\backslash \{s\}; \mathfrak{L},D)$.
\EndFor
\State $s^* = \arg \min_{s\in S} R(s)$
\State $S^* = S\backslash \{s^*\}$   \quad \quad \quad (Elimination)
\EndWhile
\State \Return $S^*$
\end{algorithmic}
\end{algorithm}

\subsubsection{Results}\label{sec:FeatureSelection:result2}

\begin{table*}
\caption{The results of multiple-feature selection based on RFE. The values on the left/right of each table cell are the results of three-feature/four-feature selection.
the performance metrics of our models were calculated with four types of sampling, including ordinary training set (noted as "NonSampling"), SMOTE training set ("SMOTE"), under-sampling training set ("UnderSample") and over-sampling training set ("OverSample").  Selected features depend on the used classifiers as well as the sampling types. Detailed analyses are made in Section \ref{sec:FeatureSelection:result2}. } \label{multiresult}

\begin{tabular}{lllllllllllllllll}
\hline
Method&Sampling&feature ID&Recall (\%)&Precision (\%)&$F_1$ Score (\%)&{FPR (\%)}\\
\hline
DT&NonSampling&3,5,10/2,3,5,10 &{95/95      }&{97/98 }&{ 96/96    }&{0.033/0.031}  \\
\rowcolor{mygray}
&SMOTE&2,5,10/2,5,10,11 &{98/96      }&{81/85  }&{89/90    }&{0.307/0.222}\\
\rowcolor{mygray1}
&UnderSample&2,5,10/2,5,10,11&{98/98      }&{51/51  }&{67/67    }&{1.260/1.260}\\
\rowcolor{mygray2}
&OverSample&1,10,14/1,4,10,14&{97/96      }&{75/84  }&{85/89   }&{0.420/0.289}\\

AdaBoost&NonSampling&3,5,9/3,5,9,16&{ 95/95     }&{  99/99  }&{ 97/97     }& {0.015/0.011}\\
\rowcolor{mygray}
&SMOTE&3,5,9/3,5,9,10&{ 99/99     }&{  78/84  }&{ 87/91    }& {0.360/0.258} \\
\rowcolor{mygray1}
&UnderSample&3,5,6/3,5,6,10&{  99/99     }&{  47/60  }&{ 64/75   }& {1.491/0.876 } \\
\rowcolor{mygray2}
&OverSample&3,5,9/3,5,9,16 &{  99/99     }&{  73/74  }&{ 84/85   }& {0.493/0.460}\\

XGBoost&NonSampling&3,5,10/3,5,810&{ 97/97      }&{ 97/97  }&{ 97/97   }& {0.033/0.033}\\
\rowcolor{mygray}
&SMOTE&3,5,10/2,3,5,10 &{ 96/\textbf{99}     }&{  47/88  }&{ 63/93  }&  {1.449/\textbf{0.169}  }\\
\rowcolor{mygray1}
&UnderSample&3,5,10/2,3,5,10&{ 99/99     }&{   60/61  }&{ 75/75   }& {0.878/0.849}\\
\rowcolor{mygray2}
&OverSample&3,5,10/3,5,7,10&{ 98/98     }&{  95/94  }&{ 96/96     }& {0.073/0.078}\\

GBoost&NonSampling&3,8,10/3,8,10,11&{ 96/96      }&{ 98/97 }&{  97/97    }& {0.031/0.040} \\
\rowcolor{mygray}
&SMOTE&3,5,10/3,5,10,11&{ \textbf{99}/\textbf{99} }&{ 89/93  }&{ 94/95 }&{\textbf{0.162}/\textbf{0.102}}\\
\rowcolor{mygray1}
&UnderSample&5,8,10/1,5,8,10&{ 98/98      }&{ 53/53  }&{ 69/69    }& {1.158/1.158}\\
\rowcolor{mygray2}
&OverSample&3,5,10/3,5,10,14&{98/97      }&{ 95/96  }&{ 96/97   }& {0.071 /0.031}\\

RF&NonSampling & 1,9,10,15 &{ 94   }&{    98 }&{ 96}&    {0.031}  \\
\rowcolor{mygray}
&SMOTE & 1,5,8,10 &{ \textbf{98}     }&{  76  }&{85   }&{\textbf{0.420}} \\
\rowcolor{mygray1}
&UnderSample & 1,5,8,10 &{ 98    }&{   59 }&{ 73  }& {0.913}\\
\rowcolor{mygray2}
&OverSample&1,5,8,10 &{ 96    }&{   80  }&{87  }& {0.318}\\

\hline
\end{tabular}
\end{table*}

The procedures of the experiments were the same as Section \ref{sec:FeatureSelection:result1} except for the feature selection step.
We evaluated the performance of our models with three and four features selected by RFE, respectively. Since three features are too limited for RF, we only trained it in the case of four features. The results were shown in Table \ref{multiresult}.
Conclusions were drawn as follows.

\begin{figure}
  \centering
  \includegraphics[width=8cm]{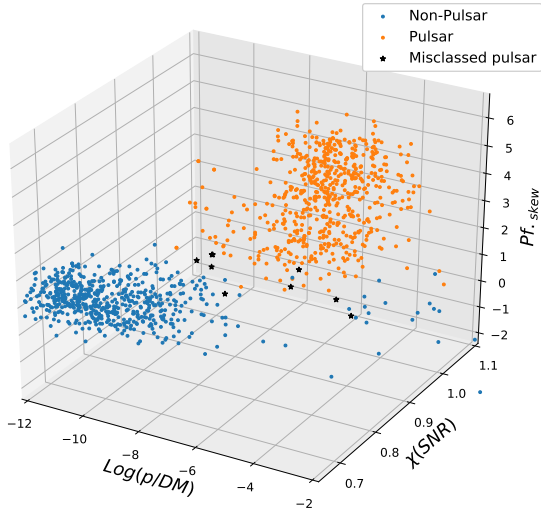}\\
  \caption{The distribution of pulsars and non-pulsars in the space of  features with ID 3,5,10. It shows great separability between pulsars and non-pulsars on these three features. However, there is a little overlap . All of the misclassified pulsars are found in this overlap area.}\label{3_5_10}
\end{figure}

\begin{itemize}
\item \textit{{The influence of class imbalance.}} As before, the models trained using the balanced sets show better performance on recall but a little worse on precision.  Considering both the recall rate and the FPR, models trained by SMOTE achieved excellent performance (a high recall $\sim$ {99\%} and a lower FPR) comparing with the other balanced samplings approaches. For example, GBoost (whose inputs are ID 3,5,10,11) trained by SMOTE dataset achieved a recall rate of 99\% and an FPR of 0.102\%.

\item \textit{Performance of the models.} Compared with the results of univariate-feature and double-feature selection, the multiple-features selection shows better performance on recall rate, precision, $F_1$ score and FPR. Firstly, the recall rate (98\%-99\%) is larger especially for models trained on the balanced sets. Secondly, it shows greater precision ($96\%$-$99\%$) for models trained in the unbalanced sets. Even so, we concerned more  about recall rather than precision. As for FPR, the models of four features computed by SMOTE data sets achieved some performances as low as, while the models trained on the unbalanced data are lower, even less than 0.03\%. Experimental results show that, only using three or four features, models such as Adaboost, Gboost, XGboost can achieve satisfactory performances.

\item \textit{Selected features.} The selected features by RFE {varies between} different models. By statistically analyzing, the features with ID 3,5,10 are the most frequently used. Results from RFE algorithm show that the selected features of each model include at least one of them. In particular, with only these three features, our GBoost achieved the performance of 99\% for the recall rate and 0.162\% for FPR. We plotted the distribution of pulsars and non-pulsars on the features with ID 3,5,10, which represents $Log(p/DM)$, $\chi_{SNR}$ and $Pf._{skew}$, correspondingly (Figure \ref{3_5_10}).
    As we have seen, it shows great separability between pulsars and non-pulsars on these three features. That is to say, a candidate to be a pulsar tends to show a small $Log(p/DM)$, a large $\chi_{SNR}$ and a positive $Pf._{skew}$. On the other hand, the features extracted from the acceleration curve (ID 15-18) provided minimal contribution in our trials. {There are very few binary pulsars in the dataset and thus features from the acceleration curve are not contributing much to the classification process.}

\item \textit{DT vs. ensemble algorithm.} The performance of the ensemble algorithms is better than DT, which is consistent with our previous analysis on ensemble algorithms. Both boosting and bagging algorithms are effective for pulsar detection.

\item \textit{Misclassified pulsars.}  As before, the pulsars frequently mispredicted  in our models were plotted, taking $Log (p/DM)$ (ID 3), $\chi_{SNR}$ (ID 5) and $Pf._{skew}$ (ID 10) as coordinates (Figure \ref{3_5_10}). It is found that there is a little overlap between the pulsars and non-pulsars where all of the misclassified pulsars are found.

\end{itemize}

\subsection{{Computational Complexity}}\label{sec:ComputationComlexity}
The GS algorithm  has the potential for finding a globally optimal solution to the problem, while the RFE is a greedy searching method which might be caught in a local optimum. However, compared to the RFE algorithm, the GS algorithm has more computational complexity \citep{Papadimitriou 2003}. Denote $m$ as the the number of features in the training dataset, $n$ as the sample number. Then the computational cost for a learner $\mathfrak{L}$ such as DT can be expressed as $T_{\mathfrak{L}}(m,n)$, an increasing function of both $m$ and $n$. If $k$ is the number of selected features ($k<m$), the computational cost of GS is $\binom{m}{k}T_{\mathfrak{L}}(k,n)$, while the computational cost of REF is $\sum_{i=k+1}^{m}iT_{\mathfrak{L}}(i,n)$. Therefore, the computational cost depends on $m$, $k$ and $T_{\mathfrak{L}}$. Take $k=8$, $m=18$ and $\mathfrak{L} = DT$ for example, the GS will consume about 200 times as much time as RFE.

\subsection{{Experimental Process}}\label{sec:FeatureSelection:implemantation}

The experiments were conducted with the following four procedures.
First, we split the data into training data set (50\%) and test data set (50\%), and balanced the training data by means of under-sampling, over-sampling and SMOTE (Table \ref{sample}). Both the training data and test data are normalized before being input into the proposed models.
Then, feature selection was made. We trained our models (Table \ref{ourmodel}) based on our proposed algorithms (GS and RFE), and computed the selected features in this step.

Next, the hyper-parameters of the ensemble learning algorithms should be established to obtain reliable and stable models. There are two primary hyper-parameters in these models, i.e., the maximum depth of the DTs of pre-pruning, the number of learners. Here, DT as the base learner of our models, is a classification and regression tree (CART) which is based on the "Gini coefficient" \citep{Breiman 1984}.
To get these hyper-parameters, a method based on grid searching was adopted and five folded cross-validation were implemented on the training data. The results were aggregated in Table \ref{parameter}.

\begin{table}
        \begin{center}
        \caption {The hyper-parameters {of the models in our work}. }\label{parameter}
        \begin{tabular}{lllllll}
        \hline \noalign{\smallskip}
          Model & Learner &Hyper-parameters\\
        \hline\noalign{\smallskip}
        DT &  & max\_depth = None, min\_leaf=10 \\
        AdaBoost & DT & max\_depth=1, n\_learners=250\\
        XGBoost & DT & max\_depth=3, n\_learners=500\\
        GBoost & DT & max\_depth=5, n\_learners=250\\
        RF & DT & max\_depth=12, n\_learners=1250\\
         \hline\noalign{\smallskip}
        \end{tabular}
        \end{center}
\end{table}

 Finally, the test data set was used to evaluate the models. The inputs into these models were the selected features (Table \ref{single_double_result} and \ref{multiresult}), while the hyper-parameters of these models were searched out in Table \ref{parameter}.

In this section, the GS was used for univariate-feature selection and double-feature selection, while the RFE was implemented for multiple-feature selection. Tables \ref{single_double_result} and \ref{multiresult} show the selected features of our models.
Using few features, the models {produced using boosting algorithms showed} good performance measurements.
Besides, it is found that the SMOTE technique was helpful for solving imbalance problem.

\section{Discussion}

%In this section, we first try to add more features to the models and discuss the their performance. Then,
%analysis is made to explain why few Features work in our models.
\subsection{Exploration on Adding More Features}

We explored the idea of adding extra features to the aforementioned models for further improving their performance.
Therefore, experiments were designed to investigate the relationship between the number of features for models and the performances of the models.
We  calculated the recall rate and FPR of all the models. These models were trained using the SMOTE dataset, and their features were  selected based on the RFE. The result shows that the recall of each model increases and the FPR of each model decreases at the beginning (Fig. \ref{recall_FPR}). However, as the the number of features increases, each model achieves a stable and high recall rate as well as a stable and low FPR.
When both the recall rate and FPR of a model are stable simultaneously, we can obtain the optimized number of features.
Therefore, the optimized number of features is 4 for Adaboost and Gboost, 5 for DT and 6 for RF and XGboost.
Also, it is found found that the models of ensemble learning (whose recall is more than {98\%}) are better than DT (whose recall is $\sim${97\%}), while the models of boosting (whose recall is $\sim${99\%}) performed better than the models of bagging (whose recall is $\sim${98\%}) in our experiments on HTRU-S (Fig. \ref{recall_FPR}).
At the same time, it is noticed that the performance of DT seems to get worse as the number of features increasing (Fig. \ref{recall_FPR}).
The reason is that a single DT classifier may study more details and split up to be a large tree when more information from extra features are added. This tree fits the training data very well but performs a little worse in the test set, which is known as \textit{over-fitting}. This is a case illustrating {the robustness of the ensemble learning methods towards the over-fitting problem compared to a single DT classifier}.

\begin{figure}
  \centering
  \includegraphics[width=8cm]{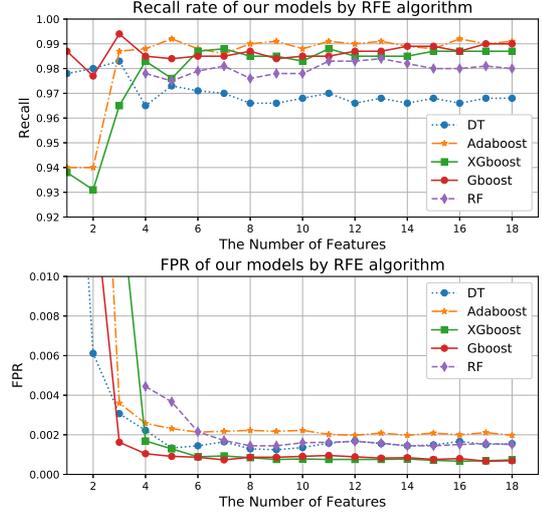}\\
  \caption{ The relationship between the number of features in the models and their performances.
  All of  the models were trained using the  SMOTE data set and their features increased based on RFE algorithm.
  It shows that each model achieves a stable and high recall as well as a stable and low FPR when the number of features is large enough. To be stable, the optimized number of features can be 4 for Adaboost and Gboost, 5 for DT and 6 for RF  and XGboost.} \label{recall_FPR}
\end{figure}

\subsection{ {Benefit of Few Features}}

A model with very few features to separate pulsar candidates is of significance. For one thing, the relevant feature can be interpretable.
The feature selection is a process of dimensionality reduction without any transformation of data. Thus, each feature completely keeps its information with its physical meaning or statistical description which can be explained straightforwardly. As we have seen, the result of double-feature selection showed great relevance for two features: SNR and $Pf_{skew}$ (Figure \ref{1to10}), while the result of triple-feature selection was $Log(P/DM)$, $\chi_{(SNR)}$ and $Pf_{skew}$ {for GBoost model (Figure \ref{3_5_10}).
Among the selected features, $Pf._{skew}$ was frequently chosen as the most distinguishing feature.

The $Pf._{skew}$ is high for most pulsars. The reason comes from the fact that most pulsar signals have small duty cycles. In other words, the profile of a pulsar signal in HTRU-S generally has a small on-pulse region and a large off-pulse region. The amplitudes in the off-pulse region are much smaller than those in its on-pulse region.
As a result, the skewness of the folded profile is large for most pulsar signals. By contrast, the $Pf._{skew}$ values are much lower for noises than those of pulsar signals. In fact, noise signals have smaller off-pulse region and larger on-pulse region, which makes the distribution of their amplitudes less skewed.}

%\begin{figure}
%  \centering
%  \includegraphics[width=4.6cm]{psrDM.eps}\\
%  \caption{ An explanation on a positive $Pf._{skew}$ for pulsars. The scattering of the signals often results in a smear tail to the right (especially sensitive to signal with low frequency). This smear tail to the right causes a positive skewness of folded profile (PSR B0833-45 in Vela; Ables,et al. 1973).} \label{scatter}
%\end{figure}

For another, models with very few features may result in better performance. To compare the performances, the metrics of our models with 2-4 features as well as those of the existing automated candidate classifiers on HTRU-S were shown (Table \ref{select-result}). We achieved {99\%} on recall and 0.162\% on FPR with only three features, while {99\%} on recall and 0.102\% on FPR with only four features. Compared with these existing results on HTRU-S, our performance was improved with fewer features.

\begin{table*}
\caption{ Comparison of performance with the existing automated candidate classifiers on HTRU-Medlat. Compared with these existing results on HTRU-S, the performance was improved with fewer features.
} \label{select-result}

\begin{tabular}{lclllllll}
\hline
& {Algor.}& Feature Num &  Sampling  &Classifier & Features {ID}  & Recall (\%)  &FPR (\%) \\
\hline
{Our Work} &{GS}  & {2}  & {SMOTE} & {Adaboost} &  {1,10} & {99}   & {0.651} \\
& {RFE}& \textbf{3}&SMOTE & GBoost &\textbf{3,5,10}&\textbf{{99}}    &  \textbf{0.162}  \\
\noalign{\smallskip}
&{RFE} & \textbf{4}&SMOTE & GBoost &\textbf{3,5,10,11}&\textbf{{99}}&  \textbf{0.102 }  \\

\citet{Bates:2012} &{/}  & 22 & NonSampling & ANN&    /  & {85}  &  1 \\

\citet{Morello 2014} &{/}  & 6 &  NonSampling        & SPINN    &    1,2,3,4,5,6  & {98}/{99}/{100}  &  0.01/0.11/0.64 \\
\citet{lyon 2016} &{/}  & 8      & Data Stream               & GHVFDT    &    7,8,9,10,11,12,13,14  & {92.8}  &  0.5 \\

\hline
\end{tabular}
\end{table*}

\subsection{{Discussion on Feature Candidates}}

In the previous sections of this work, we combined both empirical features and statistical features into our feature candidates to achieve better performance. However,
 to show the relationship between the performance and the number of features which are only from
 empirical features or statistical features, we implemented RFE algorithms in these two cases in this section. Unlike before, candidate features were only the empirical features from \cite{Morello 2014} in the first case and only statistical features from \cite{lyon 2016} in the other case. The results were shown in Fig. \ref{recall_FPR2}

In the first case, as the number of features increases, the recall rate increases and the FPR decreases until the number of features is greater than 4. The recall is high ($\sim$ {99\%}) for all the ensemble learning models but a bit lower for DT. However, the FPR is as large as about 0.5\% even if all the empirical features were used. By contrast, in the other case, both the recall and FPR tend to be stable when the number of features is greater than 2 for all the models. This means that the extra statistical features can hardly improve the performance. In other words, only very few statistical features are useful for our models and the others may be redundancy or irrelevance. Furthermore, the FPR is higher than what we obtained in Section 4 in both cases. Thus, combining different types of features and choosing from the candidate pool play a crucial role in decreasing the FPR.

\begin{figure}
  \centering
  \includegraphics[width=8cm]{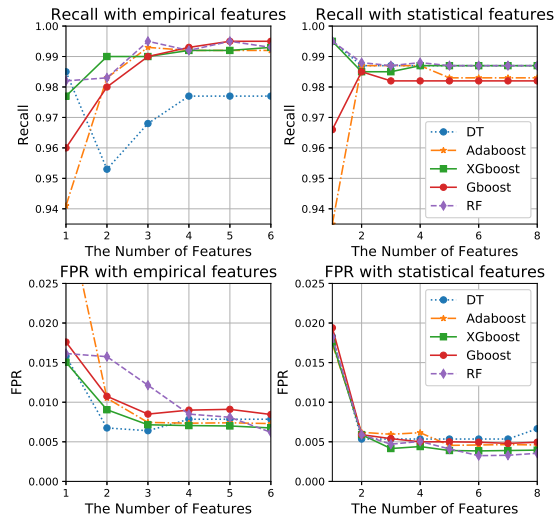}\\
  \caption{The recall rate and FPR of the models as the features increase. On the left, the features are selected only from empirical features (ID 1-6); On the right, the features are selected only from statistical features (ID 7-14). All of the models were trained  using the SMOTE data set and their features increase based on RFE algorithm.} \label{recall_FPR2}
\end{figure}

\subsection{{Misclassified Pulsars}}
Despite the success of our work, we still pay attention to the misclassified pulsars.
These misclassified pulsars were frequently misjudged
by our models after multiple-feature selection in the test data set.
It was suspected that the models with very few features may miss out on a good feature set , while the models with more feature features might be able to recognize them correctly. However, we subsequently verified through a series of tests by training models with all combinations of the 18 features, and found that adding or changing features of the models did not overcome these misclassification problems.

\begin{figure}
  \centering
  \includegraphics[width=8cm]{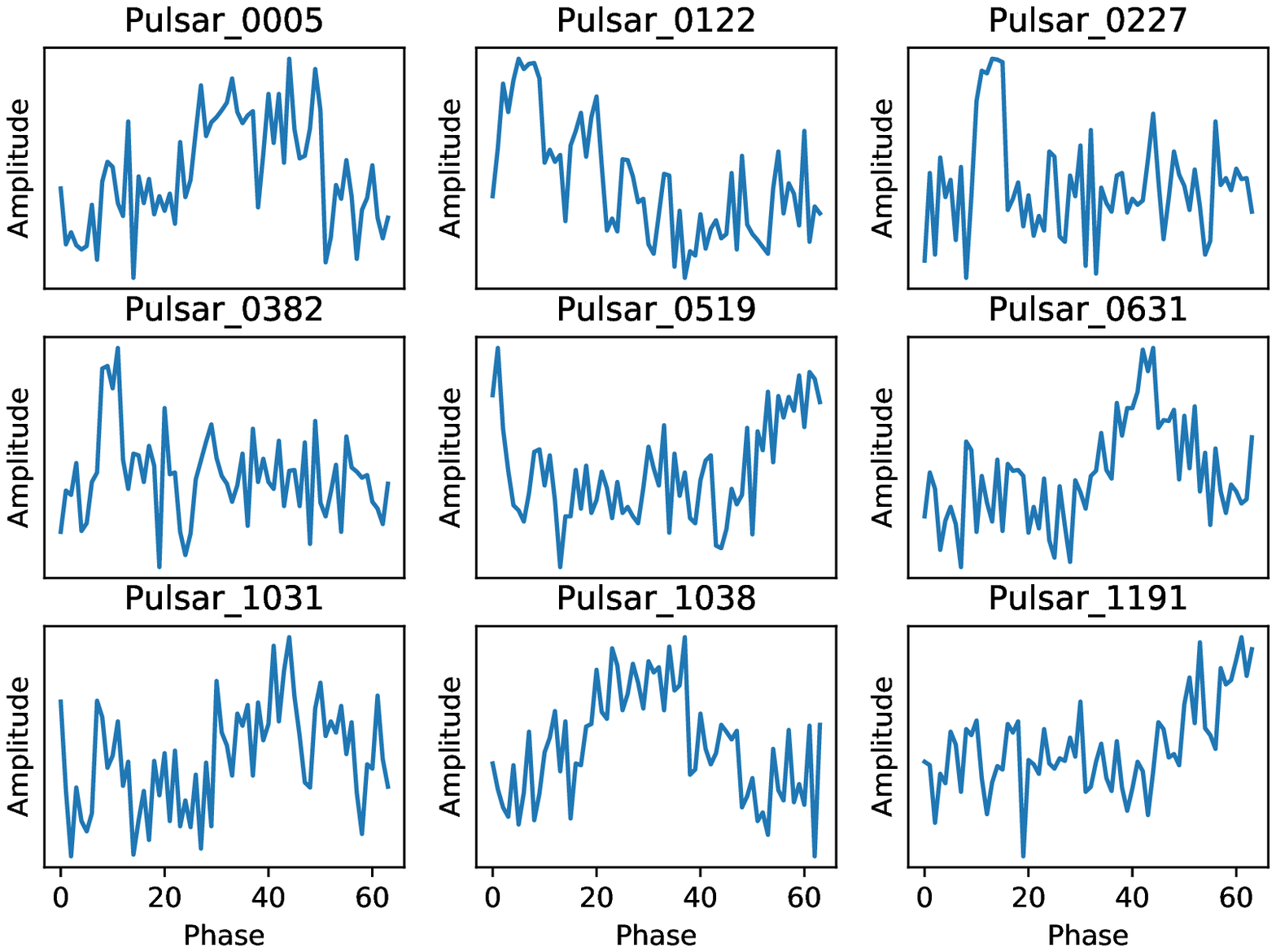}\\
  \includegraphics[width=8cm]{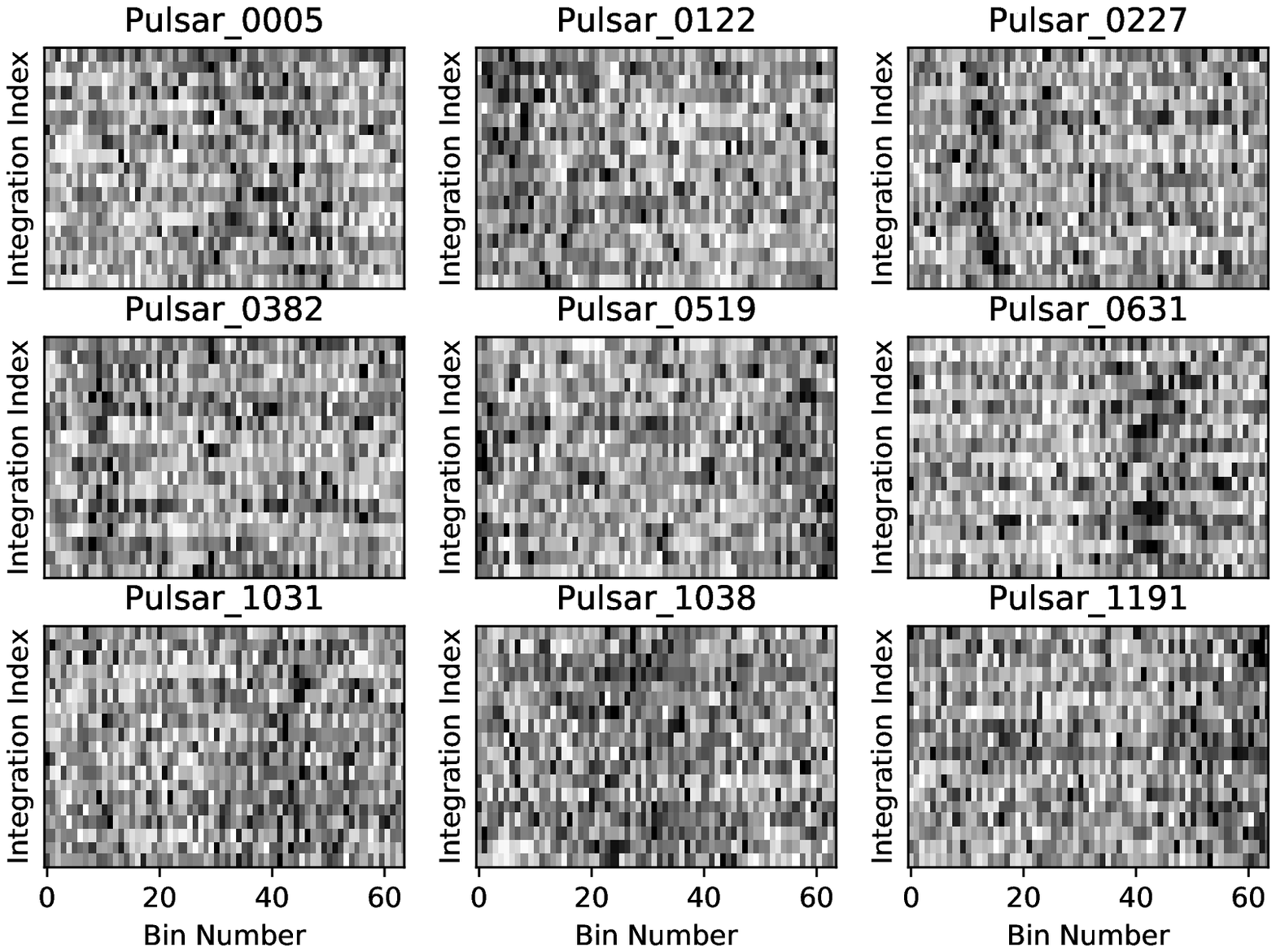}\\
  \caption{The folded profiles plots and sub-integrations plots of nine misclassified pulsars. In the experiment, these misclassified pulsars were misjudged by at least two different models. The plots on the top show that the folded profiles of these pulsars are heavily scattered, while the plots on the bottom shows that their corresponding signals in time domain are so weak that they can hardly be recognized. } \label{Missed}
\end{figure}

To further study the characteristics of these false-negative pulsars, we have shown both their folded profiles and their time-domain plots of the sub-integration (see Fig. \ref{Missed}). The figures show that the folded profiles of these pulsars (on the top) are heavily scattered and these signals in their time-domain plots (on the bottom) are so weak that they can hardly be recognized correctly by our naked eyes.
It was suspected from Fig. \ref{Missed} that these signals may have large duty cycle.
Thus, the scatter plot of $log(\texttt{SNR})$ vs. $D_{eq}$ was given (see Fig. \ref{Cause}). We found that these false negative pulsars were mainly distributed in the overlapping region of pulsars and non-pulsars, and can be characterized by:
 \begin{itemize}
 \item{
Larger $D_{eq}$. For most of the pulsars in HTRU-S, the $D_{eq}$ is lower than 5\%, while the $D_{eq}$ of some misclassified pulsars is larger than 10\%. Pulsar\_0519 in Fig. \ref{Missed} belongs to this type.}
\item{
Lower SNR. The SNR of these misclassified pulsars is as low as the non-pulsars, even less than 12.
It seems that a few misclassified pulsars with large SNR (about 12) in Fig. \ref{Cause} should be classified correctly by SNR-$D_{eq}$ space. However, we found that their $Pf._{skew}$s are as low as those from non-pulsar signals ($Pf._{skew}$ is one of the most relevant features for candidate selection), and therefore
they failed to be recognized by some ML models.
}
\end{itemize}
Further investigation in other features found that these mislabeled pulsars are almost on the boundary of pulsars and non-pulsars.
Thus, to raise the recall rate of the ML, one can create new features just like what has been done in \cite{Tan 2017}, or improve data quality from the survey.

\begin{figure}
  \centering
  \includegraphics[width=8cm]{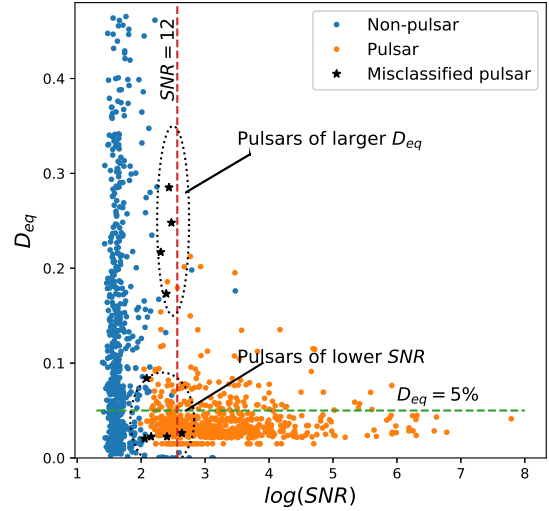}\\
  \caption{The distribution of the misclassified pulsars in our work. They can fall into two main categories: pulsars with larger $D_{eq}$ and pulsars with lower SNR.} \label{Cause}
\end{figure}

\section{Conclusions}

In this paper, the inputs into an ML {classifier for pulsar candidates} were investigated.
We attempted to improve the performances of the ML models by using as few as possible features.
We, therefore, creatively used feature selection algorithms of wrapped methods, resulting in very relevant features with their numbers from one to four. Two kinds of algorithms of wrapped methods were proposed: Grid Search (GS) and Recursive Feature Elimination (RFE).
Several techniques were applied in this process, including three sampling methods  to balance dataset, ML models of ensemble learning, a combination of candidate feature set consisting of six empirical features and twelve statistical features, and two recall-driven feature selection algorithms.
The results of optimal selected features and their performance measures have been schemed (Table \ref{single_double_result}-\ref{multiresult}).

We come to some meaningful results based on the feature selection. First of all,
pulsar candidates can be classified efficiently by machine learning models with very few features as inputs. For instance, only two features of $log(\text{SNR})$ (ID 1) and $Pf._{skew}$ can separate most of the pulsars from non-pulsars by AdaBoost {with a recall of 99\%} and an FPR of 0.548\%. Only three features including $log(P/DM)$ (ID 3), $\chi_{ (SNR)}$ (ID 5) and $Pf._{skew}$ (ID 10) can model a GBoost, achieving a recall of 99\% and an FPR of 0.162\%. The GBoost with four features results in less FPR (0.102\%). Secondly, the number of features in the models has been discussed and come to the conclusion that 4 features are enough for models of boosting (Adaboost, Gboost and XGboost), while 6 features are required for RT and 5 for DT to achieve some optimized} performance measures. Furthermore, it is found that FPR is a bit higher ($\sim$ 0.5\%) if we select features only from either empirical or statistical features. In other words, the combination of empirical features and statistical features as a feature candidate pool plays a significant role in obtaining a lower FPR. In addition, SMOTE is a useful tool in pulsar detection, as the models trained by SMOTE dataset achieved a better recall rate.
Finally, the misclassified pulsars in our models were  investigated,  and it is found that they were either weak signals or had large duty cycles in folded profile.

The combination of feature selection algorithms, SMOTE sampling method and ensemble learning models has been verified valid in our work.
 They together provided a key point to further application of machine learning in pulsar candidate selection, which goes some way towards improving the performance of high recall and low FPR when confronting high candidate numbers coming from next-generation radio telescopes such as the SKA.

\section*{Acknowledgements}
      Authors are grateful for the suggestions from Youling Yue of FAST group of National Astronomical Observatories, Chinese Academy of Sciences, and for supportings from the National Natural Science Foundation of China (grant No: 11973022%, 61273248, 61075033, 11403056
      ), and the Joint Research Fund in Astronomy (U1531242) under cooperative agreement between the National Natural Science Foundation of China (NSFC) and Chinese Academy of Sciences (CAS), and the Major projects of the joint fund of Guangdong and the National Natural Science Foundation (U1811464) to our free explorations, the Natural Science Foundation of Guangdong Province (2014A030313425, S2011010003348), China Scholarship Council (201706755006).

%%%%%%%%%%%%%%%%%%%%%%%%%%%%%%%%%%%%%%%%%%%%%%%%%%%
%
%%%%%%%%%%%%%%%%%%%%% REFERENCES %%%%%%%%%%%%%%%%%%
%
%% The best way to enter references is to use BibTeX:
%
%%\bibliographystyle{mnras}
%%\bibliography{example} % if your bibtex file is called example.bib
%

%% Alternatively you could enter them by hand, like this:
%% This method is tedious and prone to error if you have lots of references

%%%%%%%%%%%%%%%%%%%%%%%%%%%%%%%%%%%%%%%%%%%%%%%%%%

%%%%%%%%%%%%%%%%% APPENDICES %%%%%%%%%%%%%%%%%%%%%

%%%%%%%%%%%%%%%%%%%%%%%%%%%%%%%%%%%%%%%%%%%%%%%%%%

% Don't change these lines
%\bsp	% typesetting comment
\label{lastpage}
\end{document}